\documentclass[12pt,article]{aastex}

\def\BmV0{\mbox{(B-V)$^{\rm o}$}}
\def\VmK0{\mbox{(V-K)$^{\rm o}$}}
\def\MV0{\mbox{M$_{\rm V}^{\rm o}$}}
\def\MV{\mbox{M$_{\rm V}$}}

\def\etal{\mbox{\it{et al.}\ }}

\def\cs22892{{\rm CS~22892-052}}
\def\asec {^{\prime\prime}}
\def\amin {^{\prime}}
\def\fsec  {{\rlap.}^{\rm s}}
\def\fasec {{\rlap.}^{\prime \prime}\hskip0.05em}
\def\deg  {^\circ}

\begin{document}

\title{THE CONTINUING RADIO EVOLUTION  OF SN~1970G}

\author{Christopher J. Stockdale\altaffilmark{1}, W. M.  Goss\altaffilmark{2}, 
John J. Cowan\altaffilmark{1}, and R. A.  Sramek\altaffilmark{2}
}


\altaffiltext{1}{Department of Physics and Astronomy,
440 West Brooks Room 131, University of Oklahoma, Norman, OK 73019; 
stockdal@mail.nhn.ou.edu, cowan@mail.nhn.ou.edu}

\altaffiltext{2}{National Radio Astronomy Observatory, 
PO Box 0, 1003 Lopezville Road, Socorro, NM 87801-0387; mgoss@nrao.edu, 
dsramek@nrao.edu}

\begin{abstract}

Using the Very Large Array, 
we have detected radio emission from the site of SN~1970G 
in the Sc galaxy M101.  These observations are
 31 years after the supernova event, making SN~1970G the longest monitored
radio supernova.  
With  flux densities
of $0.12\pm 0.020$ mJy at 6~cm and 
$0.16\pm 0.015$ mJy at 20~cm, the spectral index of $-0.24\pm 0.20$
appears to have flattened somewhat  when compared   
with the previously  reported  value
of $-0.56\pm 0.11$, taken in 1990.
The radio emission at 20~cm has decayed since the 1990
observations with a power-law index of  
$\beta _{20cm} = -0.28\pm 0.13$. 
We discuss the radio properties of this source
and compare them to those of other Type II radio supernovae. 

\end{abstract}


\section{Introduction}

SN 1970G, a Type II supernova (SN) in M101 (Lovas 1970; Kirshner \etal 1973), 
was the first SN to be detected in the radio 
\cite{got72,all76}.  It then faded from view \cite{wei86}
until it was recovered again in 
1990 (Cowan, Goss \& Sramek 1991; hereafter referred to as CGS).  SN~1970G is
one of a very rare group of SNe that have been recovered
in the radio more than a decade following maximum optical light and is now
- 31 years after outburst - the longest monitored such object.

Intermediate-age
radio supernovae (RSNe) have been defined 
as having ages from $\sim$~10--100 years old (Cowan \& Branch 1985),  
spanning the period well
after the optical emission fades (usually about 2 years) and before
the turn-on of radio emission from the supernova remnant (SNR).
Chevalier (1984) proposes
 that synchrotron radiation is produced in the region of
interaction between the ejected supernova shell and the circumstellar 
shell that originated from the prior mass loss of the progenitor star. 
In such models the radio emission drops as the expanding shock wave 
propagates outward through the surrounding and decreasingly dense 
circumstellar material (CSM).  
Radio emission from SNRs is normally observed
long after the supernova phase.  Such factors as the density of the
local interstellar medium affect the turn-on time in models such as
those of Cowsik \& Sarkar (1984) based upon the Gull (1973) piston
model.  These models typically suggest a minimum of 100 years for the
formation and brightening of an SNR. 
The intermediate-age time period  
is, therefore, {\it critical} in understanding the later
stages of stellar evolution. In particular, the circumstellar 
mass-loss rate for the supernova progenitors is a crucial component in the
initiation and duration of radio emission in  models, such as those
of Chevalier (1984).

To study this transition period from SN to SNR, we and others have attempted to
detect radio emission from intermediate-age SNe. While there
have been a number of unsuccessful searches,
a few supernovae
have been detected in the radio more than a decade after explosion [SNe 1923A,
1950B \& 1957D in M83 \cite{cow85,cow94,eck98}, SN~1961V in NGC~1058
 \cite{cow88,sto01}, SN 1968D in NGC 6946 \cite{hym95}, and SN~1978K in 
NGC~1313 \cite{ryd93}].
One of the motivations for these studies has been that 
individual decades-old supernovae,  
such as SN 1970G, can undergo abrupt and rapid changes in flux 
densities on a timescale of only a few years.
Such variability was seen in the radio emission 
from SN~1980K,  which abruptly dropped after ten years of slow decline
\cite{wei92,mon98}. 
Also  the radio
luminosity from SN~1979C,  after declining steadily for years, has flattened
and become 
relatively constant, perhaps as a result of the 
supernova shock wave hitting a denser region of the CSM 
\cite{mon00}.
Observations also indicate that SN~1957D, 
after being relatively constant ($\sim$ 3 mJy), suddenly (in two years)  
faded below ($<$ 1 mJy) the level of an associated H II region
\cite{cow94}. 

In this paper,  
we report new radio observations of SN~1970G,   its
current flux density and spectral index,
and examine how its radio emission has varied during 
the time it has been monitored.

\section{Observations and Results}

The new Very Large Array\footnote{The National Radio Astronomy 
Observatory is a facility of the National Science Foundation operated
under cooperative agreement by Associated Universities, Inc.} (VLA)
 data on SN~1970G were taken from six observing runs.
In the first three, SN~1970G was observed for 4.5 total hours 
on 4 and 6 November and 3 December 2000, at 20~cm (1.425 GHz) using the 
VLA's most extended (A) configuration, 
with a maximum baseline of 34 km.
During the second group of observing runs, SN~1970G was
observed on 10 January 2001 with the
VLA in its A configuration and on 5 and 21 February 2001 with
VLA in its BnA configuration (maximum baseline of 24 km)
at 6cm (4.860 GHz) for a total of 6.5 hours.   These observations were done using the standard VLA continuum mode, 
obtaining a total of 100 MHz bandwidth in each of
the two orthogonal circular polarizations. The phase calibrator was 
J1400+621 and 3C286 was used to set the flux density scale. 
In all observations the pointing center was $14^{h}03^{m}00^{s}$ and
$+54\deg 14\amin 33\asec$ (J2000), and flux densities for 3C286 were taken from 
Perley, Butler, \& Zijlstra (2001).

Data were Fourier transformed and deconvolved using the CLEAN algorithm
as implemented in the AIPS routine IMAGR, with the
Brigg's robustness parameter of $0$, which minimizes the 
point-spread function while maximizing sensitivity.
The 20~cm and 6~cm observations of the region from 1990 were also
re-analyzed  using the same
data reduction procedures and inputs as were used on the current data.
The re-calculated results are presented in Table 1.

The 6~cm observations were combined using the AIPS routine DBCON, 
and a final 6~cm image was produced.  
The AIPS routine CONVL was used to correct for differences in the
beam size and shape since they were taken in 
different array configurations.  The 6~cm convolved beam size was
set to be $1\fasec 20 \times 1\fasec 20$.
  
The peak flux density and position for SN~1970G were estimated in
three ways using the AIPS routines IMFIT, JMFIT, and SLICE.  SLICE
produces a series of one dimensional plots of flux densities vs. 
position at constant right ascension and then at constant declination.  
The SLICE plots were examined for baseline offsets owing to
the nearby H~II region, NGC~5455 [centered $\simeq 5\sec$ from SN~1970G].
The results from the SLICE analysis are given in Table 1; the results 
from all three analysis methods agree to within the errors shown.

The position measured
for SN~1970G, from the new images and the re-analysis of the 20~cm observations
taken in 1990,  is $14^{h}03^{m}00\fsec 88 \pm 0\fsec 01$ and
$+54\deg 14\amin 33\fasec 1 \pm 0\fasec 2$.  
Uncertainties in the peak intensities are reported as the
rms noise from the observations.  At 20~cm, the beam size is 
$1\fasec 42 \times 1\fasec 21$, p.a. = $-39.18\deg$,
and the rms noise is 0.015 mJy beam$^{-1}$.
At 6~cm, the beam size is 
$1\fasec 20 \times 1\fasec 20$, p.a. = $0.00\deg$,
and the rms noise is 0.020 mJy beam$^{-1}$.
The results of our analyses are presented in Table 1 and 
Figure \ref{1}.  

\section{Discussion and Conclusions}

We have detected a radio source at the position of SN~1970G
at 20~cm and 6~cm, coincident, within the error limits, with the CGS position.
Our measured flux density at 20~cm indicates a reduction 
in the 20~cm flux density of only 11\% from 1990 to 2000
(see Table 1).
And, extrapolating from the 3.5 and 20 cm measurements reported by CGS, the 6
cm flux density in 1990 would have been
0.11 mJy, which is virtually identical to the current value of $0.12\pm 0.020$ 
mJy (see Table 1).  
The CGS observations indicated that the flux density of SN 1970G 
had dropped between 1974 and 1990 with 
a power-law index ($S \propto t^{\beta}$) of $\beta = -1.95 \pm 0.17$ at 20~cm.
Assuming the CGS power-law decline and spectral index from 1990,
we would expect a much lower 6cm flux density, $\simeq$ 0.050 mJy, and
20~cm flux density, $\simeq$ 0.090 mJy than we now measure.
The radio light-curve for SN~1970G between 1990 and 2000 
has flattened considerably and (based upon the 20~cm flux densities)
the power-law index is 
now  $\beta = -0.28\pm 0.13$.
We illustrate this dramatic change in slope in the radio evolution of 
SN 1970G in Figure \ref{2}; where we show 
the radio light curves of several 
intermediate-age SNe along with a few SNRs, 
plotting the time since supernova explosion versus the 
luminosity at 20~cm (i.e. the monochromatic luminosity).  

As indicated in Table 1 (and shown in Figure 1), 
our new observations indicate that SN~1970G  
is still  a (marginally) non-thermal radio source. 
The current spectral index, $\alpha$, of $-0.24\pm 0.20$ 
has flattened from the value of $-0.56\pm 0.11$ reported in 1990 by CGS,
although the error bars are rather large. 
We might expect a possible flattening of the radio spectrum 
as the emission from SN~1970G continues to fade.
Such was the case for  
SN~1957D, in M83, which had a very similar spectra index, 
$-0.23\pm 0.04$, at a comparable age to SN 1970G (Cowan \& Branch 1985). 
The thermal spectral index reported by Cowan \etal (1994) for SN~1957D
is actually due to the associated H II region that is now 
brighter than the faded supernova.
However, there has been no optical detection
of an associated H~II region coincident with SN~1970G, and thus  
it should be possible to continue to follow the time evolution of the 
flux density and the spectral index for this supernova for many years.

The current and previous values of the spectral
index, luminosity, and decay index for SN~1970G
are within a range of values reported for other intermediate-age
RSNe at similar ages and wavelengths 
(See Table 1,  Figure 2, and Stockdale \etal 2001).
It is clear, for example,  from Figure 2  that the radio luminosity of SN~1970G
at an age of $\simeq$~30 years is very close to that of  
the Type II SNe~1950B and 1957D, in M83, SN1968D, in NGC 6946, and SN~1961V, 
in NGC~1058, at the same stage in their evolution (Cowan \etal 1994;
Hyman \etal 1995; Stockdale \etal 2001).  
This correlation in luminosity also lends credence to the identification of
SNe~1950B and 1961V as Type II SNe.  In particular, there has 
been some  uncertainty in 
the optical position of SN~1950B that prevented a conclusive identification
of the supernovae with the 
radio source \cite{cow85,cow94}.  In the case of SN~1961V, 
recent optical observations have caused
some debate whether it was a supernova event or a luminous blue variable
(LBV) \cite{goo89}.   On the other hand, recent radio observations of SNe~1961V 
strongly suggest a supernova interpretation for this event \cite{sto01}.

The evolution of the radio flux density of SN~1970G 
is consistent with the current models for radio emission from SNe, 
which predict a general decline in radio luminosity with age and 
declining density of CSM.
Figure {\ref{2} illustrates that trend for a number of intermediate-age RSNe,
with a general  
flattening of the radio light curve 10-40 years after
the supernova event.  While SNe 1970G, 1961V, and 1950B
are decaying at different rates, we note that all of 
their radio light curves are noticeably 
flatter than that of SN~1957D at similar epochs. 
The differences in the  behavior of the individual SNe
(e.g. the rates at which their radio luminosities fade) 
could therefore be explained in terms of the density of the material 
encountered by the supernova shock. 
Thus, for example,  
the shocks associated with some RSNe
(e.g. SNe~1979C, 1970G, and 1961V) might be  
traveling through considerably denser CSM than
other similarly-aged RSNe (e.g., SNe~1980K and 1957D).
Consistent with this interpretation is the very rapid decline and disappearance
of the radio emissions of Type Ib RSNe, e.g. SNe~1983N and 1984L, 
which presumably have lower density  CSM than Type II SNe 
\cite{wei86,sra84,pan86}.   

One scenario which might explain
this increased density of CSM around some Type II RSNe 
could be that the progenitors underwent large-scale eruptions, 
akin to LBVs, prior to the supernova event.  
The mass loss rates during an LBV eruption
can be 10---100$\times$ larger than the typical supergiant mass-loss rate
\cite{hum94}.  The exact epoch at which this may have occurred depends
on the ejection velocities of the CSM during these events, the rate
of expansion of the supernova shock, and the density of the CSM.  
Unfortunately RSNe are too under-sampled
to make any definitive 
statements as to the exact nature of such a possible outburst or mass loss.
Clearly, additional radio monitoring  of SN~1970G, and  other 
RSNe, will be important in understanding  
the continuing evolution and nature of these 
relatively rare objects. 
  
\acknowledgments
We would like to thank an anonymous referee for constructive comments that
helped us to improve this Letter.
The research was supported in part by the 
NSF (AST-9618332 and AST-9986974 to JJC) 
and has made use of the 
NASA/IPAC Extragalactic Database (NED), which is operated by the Jet Propulsion Laboratory, 
Caltech, under contract with the National Aeronautics and Space Administration.

\clearpage

\begin{deluxetable}{lcccc}
\dummytable\label{A}
\tablenum{1}
\tablewidth{7.0in}
\tablecaption{Properties of Intermediate-Age Radio Supernovae}
\tablecolumns{5}
\tablehead{
\colhead { }   &
\colhead{SN~1970G}   &
\colhead{SN~1961V}    &
\colhead{SN~1957D}    &
\colhead{SN~1950B}    }
\startdata
Right Ascension (J2000)                        
& $14^{h}03^{m}00\fsec 88$
& $02^{h}43^{m}36\fsec 46$      
& $13^{h}37^{m}03\fsec 57$
& $13^{h}36^{m}52\fsec 88$      \\
Declination (J2000)                            
& $+54\deg 14\amin 33\fasec 1$ 
& $+37\deg 20\amin 43\fasec 2$ 
& $-29\deg 49\amin 40\fasec 7$
& $-29\deg 51\amin 55\fasec 7$  \\
\tableline
Distance (Mpc)
& 7.4
& 9.3
& 4.1
& 4.1 \\
\tableline
Most Recent 20~cm 
&  
& 
&
& \\
Flux density (mJy)
& $0.16\pm 0.015$
& $0.15\pm 0.026$               
& $1.22\pm 0.07$ 
& $0.72\pm 0.04$             \\
Supernova age (years)
&  30.29
&  37.76
&  34.04
&  41.80  \\
Luminosity (ergs s${^{-1}}$ Hz$^{-1}$) 
& 1.1 $\times 10^{25}$
& 1.5 $\times 10^{25}$
& 2.4 $\times 10^{25}$
& 1.4 $\times 10^{25}$   \\
\tableline
Early 20~cm 
& 
& 
&
& \\
Flux density (mJy) 
& $0.18\pm 0.017$
& $0.23\pm 0.020$               
& $2.7\pm 0.12$
& $0.7\pm 0.08$               \\
Supernova age  (years)
& 19.74
& 22.93
& 25.24
& 33.00   \\
Luminosity (ergs s${^{-1}}$ Hz$^{-1}$) 
& 1.2 $\times 10^{25}$
& 2.4 $\times 10^{25}$
& 5.3 $\times 10^{25}$
& 1.4 $\times 10^{25}$   \\
\tableline
Most Recent 6~cm 
&  
&
& 
& \\
Flux Density (mJy)
& $0.12\pm 0.020$
& $0.06\pm 0.008$               
& $1.39\pm 0.04$
& $0.37\pm 0.03$               \\
Supernova age  (years)
& 30.50
& 38.12
& 32.82
& 40.58  \\
Luminosity (ergs s${^{-1}}$ Hz$^{-1}$) 
& 7.9 $\times 10^{24}$
& 6.5 $\times 10^{24}$
& 2.8 $\times 10^{25}$
& 7.4 $\times 10^{24}$  \\
\tableline
Most Recent Spectral Index\tablenotemark{a}  $\alpha$ 
& $-0.24\pm 0.20$
& $-0.79\pm 0.23$        
& $+0.11\pm 0.06$
& $-0.57\pm 0.08$                \\
Earlier Spectral Index\tablenotemark{a}  $\alpha$
& $-0.56\pm 0.11$
& $-0.44\pm 0.15$        
& $-0.23\pm 0.04$
& $-0.55\pm 0.13$
                \\
Most Recent 20~cm Decay Index\tablenotemark{b}   $\beta$
& $-0.28\pm 0.13$
& $-0.79\pm 0.23$        
& $-2.66\pm 0.07$
& $+0.12\pm 0.13$                \\
\enddata

\tablenotetext{a} {\ $S \propto \nu^{\alpha}$, measured between 20~cm and 6~cm, 
execept for the earlier SN~1970G which was measured between 20~cm and 3.5~cm }
\tablenotetext{b} {\ $S \propto t^{\beta}$}
\tablerefs{ \cite{cow91,cow94,kel96,sah95,sil96,sto01} }
\end{deluxetable}

\clearpage
\pagestyle{empty}

\begin{figure}
\epsscale{0.6}
\plotone{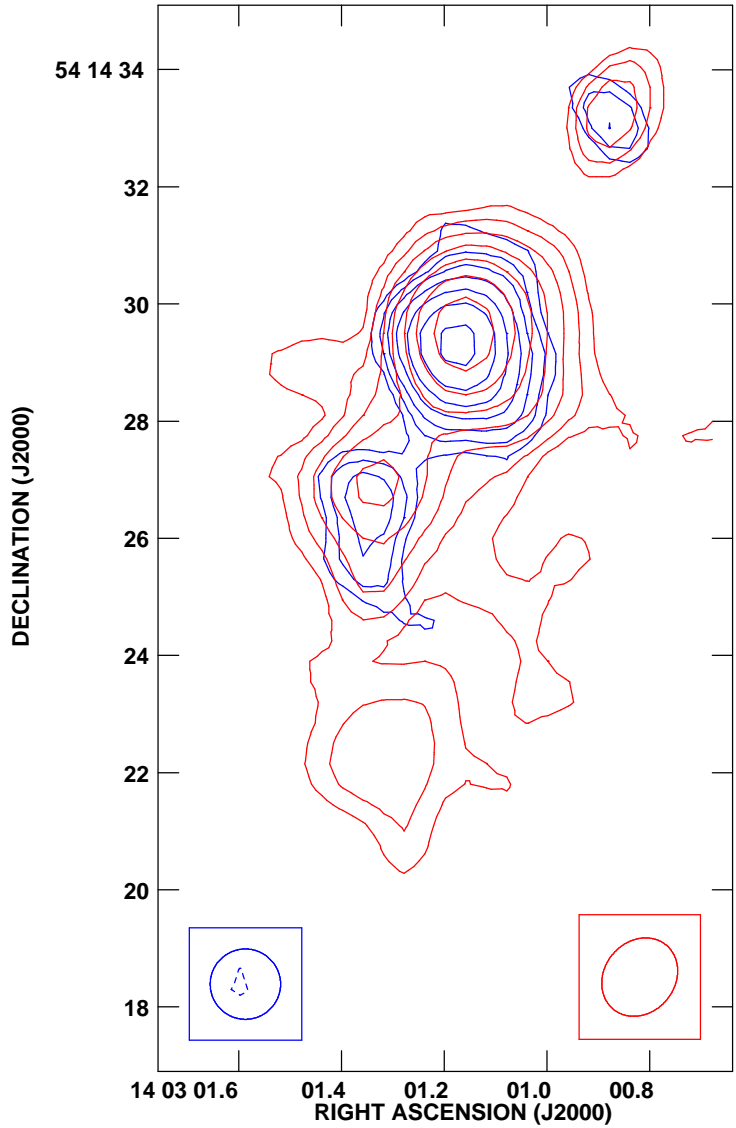}
\caption{
Radio contour images at 20~cm (in red) and 6~cm (in blue) of
SN~1970G (the source above and to the right of the central H~II region,
NGC~5455).
Contour levels at both wavelengths
are -0.060, 0.060, 0.086, 0.12, 0.17, 0.24, 0.34, 0.48, 0.68, and 0.96 mJy
beam$^{-1}$.
At 20~cm, the beam size is (shown in lower right)
$1\fasec 42 \times 1\fasec 21$, p.a. = $-39.18\deg$,
and the rms noise is 0.015 mJy beam$^{-1}$.
At 6~cm, the beam size is (shown in lower left)
$1\fasec 20 \times 1\fasec 20$, p.a. = $0.00\deg$,
and the rms noise is 0.020 mJy beam$^{-1}$.
\label{1}}
\end{figure}  

\begin{figure}
\epsscale{0.95}
\plotone{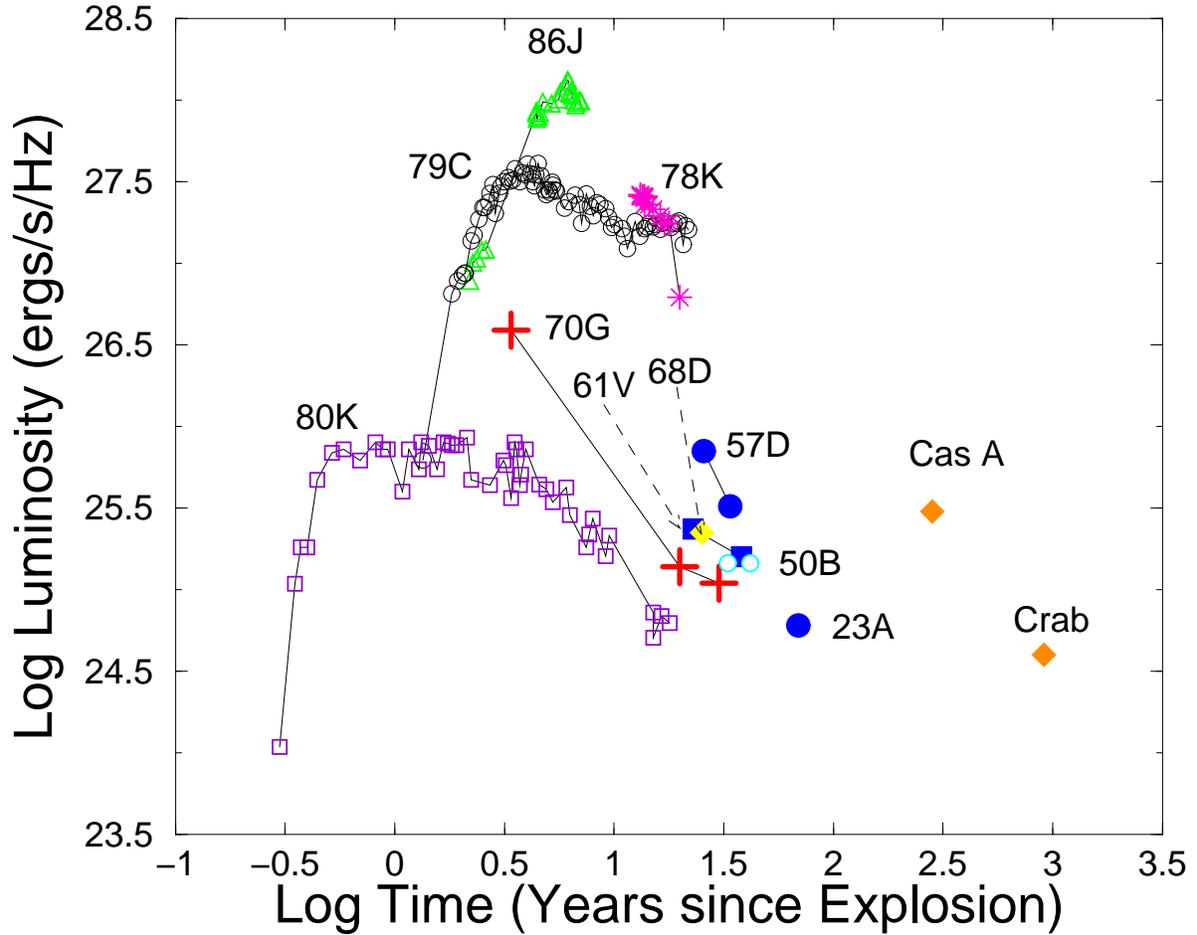}
\vskip .2truein
\caption{
Radio light curve for SN~1970G at 20~cm (indicated by red crosses) 
compared to several RSNe
and SNRs.  Data, fits and distances for SN~1923A, from Eck \etal (1998)
and Saha \etal (1995); for SNe 1950B (cyan hoops) \& 1957D, from
Cowan \etal (1994) and 
Saha \etal (1995); for SN~1961V (blue squares), Stockdale \etal (2001) and 
Silbermann \etal (1996); for SN~1968D (yellow diamond), 
from Hyman \etal (1995) and Tully (1988); 
 SN~1970G, from this paper, Cowan \etal (1991), 
and Kelson \etal (1996); for SN~1978K, from Ryder \etal (1993), 
Schlegel \etal (1999), and Tully (1988);
for SN~1979C, from Weiler \etal (1986, 1991), Montes \etal (2000), and 
Ferrarese \etal (1996); for SN~1980K, 
from Weiler \etal (1986, 1992), Montes \etal (1998), and 
Tully (1988); and for SN~1986J, from Rupen \etal (1987), 
Weiler \etal (1990), and Silbermann \etal (1996).  
Luminosities for Cas~A and the Crab from Eck \etal (1998).
\label{2}}
\end{figure} 

\end{document}